# ON NON-MEASURABLE SETS AND INVARIANT TORI

by

Piotr Pierański [+] and Krzysztof W. Wojciechowski [++]

[+] *Poznań University of Technology*
*Piotrowo 3, 60 965 Poznań, Poland*

[++] *Institute of Molecular Physics, Polish Academy of Sciences*
*M. Smoluchowskiego 17, 60 159 Poznań, Poland*

**ABSTRACT**

The question: "*How many different trajectories are there on a single invariant torus within the phase space of an integrable Hamiltonian system*?" is posed. A rigorous answer to the question is found both for the rational and the irrational tori. The relevant notion of non-measurable sets is discussed.

## I. INTRODUCTION

Irrational invariant tori play a crucial role in physics of Hamiltonian systems. In contrast to the rational tori, they prove to be to some extent resistant to the destructive action of the non-integrable perturbations and, as the KAM theorem establishes it, the measure of the set of those tori which remain intact, though obviously distorted, is non-zero at low levels of the perturbation.[1]

It is the aim of this paper to indicate a peculiar property of irrational tori: non-measurability of sets of points which initiate on them all possible (and different) trajectories. To make the considerations which follow as clear as possible, we fix our attention on the simplest nontrivial case - a Hamiltonian system with but two degrees of freedom $q_1$ and $q_2$ whose trajectories are located within a four-dimensional phase space $\Gamma$. (Generalisation for more degrees of freedom is trivial.) We assume that the system is integrable, i.e. there exist two integrals of motion $I_1$ and $I_2$ which allow one to describe any motion of the system as two independent rotations on a two-dimensional torus; see Figure 1.

$$\varphi_1(t) = \varphi_1(0) + \omega_1 t \qquad (1)$$

$$\varphi_2(t) = \varphi_2(0) + \omega_2 t \qquad (2)$$

In the most general case, the two frequencies are different on each of the tori into which the whole phase space of the system is partitioned. There are two basic types of the tori: those for which the ratio $\omega_1/\omega_2$ is rational and those for which the ratio is irrational.

Let us ask a question: How many different trajectories are there on a single: (i) rational and (ii) irrational torus ?

To observe trajectories which move on the torus $T$ in a more convenient manner, we cut it with a Poincaré section $S$. In this plane a single trajectory is seen as a sequence of points which mark all its consecutive (past and future) passages through $S$. All the points are located, of course, on the circle $C = T \cap S$. A single point from such a sequence determines the whole (and single) trajectory. There are many different (disjoint) trajectories moving on the torus. Each of them defines on $C$ a sequence of points. Choosing single points from all such sequences allows one to construct the required set of points which initiate on the torus different trajectories. Let us denote the set by $M_0$. Thus, the question we posed above can be reduced to the following one: *How big is the set $M_0$?*.

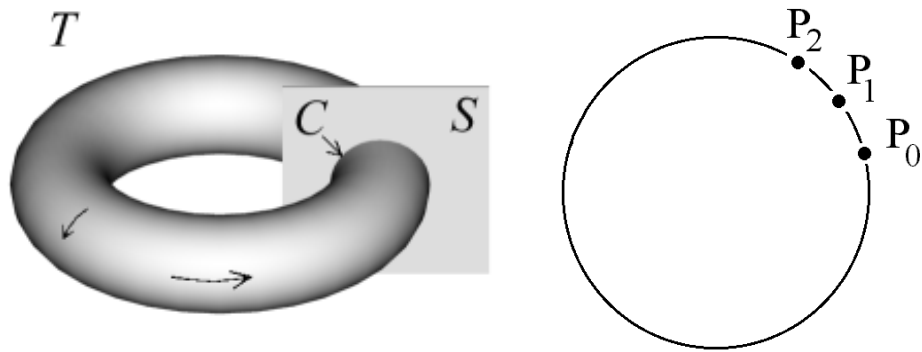

Fig. 1 Invariant torus and a Poincare section of it. A trajectory starting form point $P_0$ pierces plane $S$ in consecutive points $P_1, P_2, P_3 \ldots$ . See text.

By "how big" we mean here two things :

1. *Is the set $M_0$ countable ?*

2. *Which is its measure $\mu(M_0)$ ?*

Below we shall answer the questions, first, for the case of rational tori, then, for the case of the irrational ones.

## II. $\mu(M_0)$ ON RATIONAL TORI.

Let $T_r$ be a rational torus, i.e. a torus for which $\omega_1/\omega_2 = r = m/n$, where $m$ and $n$ are integers. Any trajectory on $T_r$ marks in $S$ as a cycle of $n$ points $\{P_k\}$, $k=0, 1, \ldots, n-1$, whose angular coordinates $\varphi_k$ are given by

$$\frac{\varphi_k}{2\pi} = \frac{\varphi_0}{2\pi} + kr \bmod 1 \qquad (4)$$

As easy to note, all trajectories which start from those points on $C$ whose $\varphi$ coordinates are located within any interval $[\varphi_0, \varphi_0+2\pi/n)$, where $\varphi_0$ is arbitrary, are different and, as such a choice is made, there are no other different trajectories.

Consequently, the set

$$M_0 = \{P \in T \cap C : \varphi(P) \in [0, \frac{2\pi}{n})\} \qquad (5)$$

can be seen as the simplest realisation of the set of points which initiate on $T$ all possible different trajectories.

Obviously, in this case, the set is uncountable and its measure :

$$\mu(M_0) = \frac{\mu(C)}{n} \qquad (6)$$

Any other choice of $M_0$ provides the same answer.

## III. $\mu(M_0)$ FOR IRRATIONAL TORI.

Let $\omega_1/\omega_2 = \rho$ be an irrational number. Now, any trajectory on $T_\rho$ is seen within the Poincaré section $S$ as an infinite, never repeating itself sequence of points $\{P_k\}$, $k = -\infty \ldots, -1, 0, 1, 2, \ldots +\infty$, whose angular coordinates are given by

$$\frac{\varphi_k}{2\pi} = \frac{\varphi_0}{2\pi} + k\rho \,(\mathrm{mod}\, 1) \qquad (7)$$

The countable set $\{P_k\}$ covers $C$ in a dense manner but is different from it: $C\backslash\{P_k\}\neq\emptyset$. Thus, there are on $C$ some points which initiate other trajectories. How to find all of them i.e. define set $M_0$? To reach the aim we shall proceed in three steps.

*Step 1*. We define within the $[0,1)$ interval a countable, everywhere dense set

$$E = \{k\rho \,\mathrm{mod}\, 1 : k \in N\} \qquad (8)$$

*Step 2*. We define in $C$ a relation $\mathfrak{R}$:

$P\mathfrak{R}Q$ if and only if there exists in $E$ an $x$ such that:

$$\frac{|P-Q|}{2\pi} = x \qquad (9)$$

In plain words the physical meaning of $\mathfrak{R}$ can be expressed as follows:

*P and Q stay in relation $\mathfrak{R}$ when they belong to the same trajectory.*

One can prove that $\mathfrak{R}$ is an equivalence relation, thus, $\mathfrak{R}$ divides $C$ into a family of equivalence classes $C/\mathfrak{R}$. In view of the physical meaning of $\mathfrak{R}$ the classes are simply Poincaré section images of trajectories which move on the torus $T$. Since the classes are disjoint, the trajectories they represent are all different. Since the classes cover all $C$ - there are no other trajectories.

*Step 3*. From each class from the family $C/\mathfrak{R}$ we take one point and put it into a set $M_0$. Obviously, the set $M_0$ can be seen as the set of points which initiate on $T$ different and all possible trajectories. Let us have a closer look at it.

First of all, we may check what happens when trajectories initiated by all points from $M_0$ pass through the Poincaré section $S$. Let sets $M_k$, $k=-\infty, \ldots, -1, 0, 1, \ldots, +\infty$, be the images of $M_0$ which appear in $C$ as the trajectories make consecutive turns on $T$. All the sets are disjoint:

$$M_k \cap M_j = \emptyset, \; i \neq j \qquad (10)$$

and their union covers whole $C$:

$$\bigcup_k M_k = C . \qquad (11)$$

Since $C$ is uncountable (continuum) and the family of sets $\{M_k\}$ is only countable, the equipollent sets $M_k$ cannot be countable. In particular, $M_0$ is uncountable. This answers the first part of the question.

Now, let us consider its second part. Since each $M_k$ can be obtained from $M_0$ by a rotation of the latter along $C$ by an angle $2\pi[k\rho \; (mod \; 1)]$, all of the sets are mutually congruent. Thus, if the measure of the set $M_0$ is $\mu(M_0)$, the measure $\mu(M_k)$ of each $M_k$ must be the same:

$$\mu(M_i) = \mu(M_j) \text{ for all } i, j \qquad (12)$$

On the other hand, in view of Eq.11, we have

$$\sum_k \mu(M_k) = \mu(C) = 1. \qquad (13)$$

*How much is* $\mu(M_0)$ ? It cannot be zero, since in that case Eq.13 would give $\mu(C)=0$, which is not true. On the other hand, it cannot be finite either, since in that case we would obtain from Eq.13 $\mu(C)=\infty$. Thus: How much is it ? We cannot say. A number which would describe the value of $\mu(M_0)$ does not exist : *the set $M_0$ is non-measurable*.

**IV. DISCUSSION**

The reasoning we described in Section III can be seen as a direct translation of the Vitali construction of a non-measurable set [2,3] onto the language of the Hamiltonian mechanics. As a careful reader may have noted, step 3 of the construction we presented makes use of the Axiom of

Choice (AC), the most controversial and at the same time, the most thoroughly studied pillar from the few ones on which the theory of sets can be built[4]. We write "can", since, being both *relatively consistent*[5] with other axioms of the set theory and *independent*[6] of them, the axiom can be used or not. Taking the former attitude, i.e. deciding to use the Axiom, one is able not only to prove a few most useful theorems (impossible to prove without the axiom), such as that *the union of countably many countable sets is countable* or that *every infinite set has a denumerable subset*, but, and this is most disturbing, a number of theorems which seem to stay in contradiction with what our common sense tells. The most famous from such theorems, proven by Banach and Tarski[7], says that *it is possible to dismount a sphere into a few (at least five) such subsets, from which, using but translations and rotations i.e. transformations which certainly preserve measure, one can mount two spheres identical with the initial one*. Obviously, the Banach-Tarski theorem stays in conflict with our common sense. But was it not like that already once with the theorems of the non-Euclidean geometry? The problem is not, we emphasise it, if the Banach-Tarski theorem is true or not; its proof is clean and completely legitimate within a certain mathematical environment. One should rather ask *if we need this branch of mathematics to describe things which happen in the world and which we study in physics laboratories*. (Non-Euclidean geometry has proven its usefulness in the description of our world at scales somewhat larger from that at which our common sense is formed.) Do we know phenomena whose description would necessarily require the use of the axiom of choice? There have been so far but a few attempts of applying the axiom of choice based mathematics to describe physical reality. The first one was the Pitowsky's work on a possible resolution of the Einstein-Podolsky-Rosen paradox via the Banach-Tarski one[8,9]. The work by Pitowsky indicated the possibility that physical paradoxes encountered in quantum mechanics can be reduced to mathematical pathologies[8]. The aim of the present note is somewhat different. It shows that such a mathematical pathology (non-measurability) appears in the formal analysis of a very simple problem of classical mechanics and cannot be avoided there; one cannot answer the question concerning the number of trajectories on irrational tori without using the axiom of choice. Consequently, if one decides to answer the question one must necessarily get in touch with the paradoxical concept of sets without measure.

A similar problem and a similar way of solving it is described in [10], where Svozil and Neufeld analyse the concept of linear chaos. A general, very vivid exposition of the problem of applicability of the set theory in the description of the physical world can be found in [11]. As Svozil argues there, the prohibition on the use of paradoxical results of the set theory cannot be accepted. Such a "No-Go" attitude, as he calls it, has no justification. According to Svozil, the No-

Go attitude should be rejected in favour of the "Go-Go" attitude, according to which results of any consistent mathematical theory may be used in the description of the physical world. From the Svozil's point of view, the present authors took a full advantage of the Go-Go attitude: using the based on the Axiom of Choice notion of non-measurable set they answered a concrete, sensible question formulated within the frames of classical mechanics. An extensive study of the links between physics and set theory was presented also by Augenstein [12], who among other examples draws our attention to the use by El Naschie of the paradoxical decomposition technique in the analysis of the Cantorian micro space-time [13].

Concluding the present work, we admit that defining the set of points which on an irrational invariant torus initiate different trajectories we did use the Axiom of Choice. From the formal point of view the definition must be seen as a nonconstructive one. In Svozil's wording: throwing out the nonconstructive bath water we would throw with it the nonmeasurable baby [10]. We think it would be not right.

**Acknowledgements**

One of the Authors (K.W.W.) is grateful to Professor William G. Hoover and to Professor Janusz Tarski for encouragement. He is also grateful to Professor M. P. Tosi and Professor Yu Lu for hospitality at the Abdus Salam International Centre for Theoretical Physics, Trieste, Italy.